%% file: main.tex
\documentclass[sigconf,natbib=true]{acmart}
\AtBeginDocument{%
  \providecommand\BibTeX{{%
    \normalfont B\kern-0.5em{\scshape i\kern-0.25em b}\kern-0.8em\TeX}}}

\copyrightyear{2023} 
\acmYear{2023} 
\setcopyright{rightsretained} 
\acmConference[SIGIR '23]{Proceedings of the 46th International ACM SIGIR Conference on Research and Development in Information Retrieval}{July 23--27, 2023}{Taipei, Taiwan}
\acmBooktitle{Proceedings of the 46th International ACM SIGIR Conference on Research and Development in Information Retrieval (SIGIR '23), July 23--27, 2023, Taipei, Taiwan}
\acmDOI{10.1145/3539618.3591724}
\acmISBN{978-1-4503-9408-6/23/07}

\usepackage{etoolbox} 
\makeatletter
\patchcmd{\maketitle}{\@copyrightpermission}{
\begin{minipage}{0.3\columnwidth}
\href{https://creativecommons.org/licenses/by/4.0/}{\includegraphics[width=0.90\textwidth]{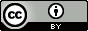}}
\end{minipage}\hfill
\begin{minipage}{0.7\columnwidth}
\href{https://creativecommons.org/licenses/by/4.0/}{This work is licensed under a Creative Commons Attribution International 4.0 License.}
\end{minipage}

\vspace{5pt}
}{}{}

\input{definitions}
\input{packages}

\begin{document}

\title{Measuring Item Global Residual Value for Fair Recommendation}


\author{Jiayin Wang}
\affiliation{%
  \institution{DCST, BNRist, Tsinghua University}
  \city{Beijing 100084}
  \country{China}}
\email{jiayin-w20@mails.tsinghua.edu.cn}

\author{Weizhi Ma}
\affiliation{%
  \institution{AIR, Tsinghua University}
  \city{Beijing 100084}
  \country{China}}
\email{mawz@tsinghua.edu.cn}

\author{Chumeng Jiang}
\affiliation{%
  \institution{DCST, BNRist, Tsinghua University}
  \city{Beijing 100084}
  \country{China}}
\email{jcm20@mails.tsinghua.edu.cn}

\author{Min Zhang*}
\affiliation{%
  \institution{DCST, BNRist, Tsinghua University}
  \city{Beijing 100084}
  \country{China}}
\email{z-m@tsinghua.edu.cn}

\author{Yuan Zhang}
\affiliation{%
  \institution{Kuaishou Inc.}
  \city{Beijing 100085}
  \country{China}}
\email{yuanz.pku@gmail.com}

\author{Biao Li}
\affiliation{%
  \institution{Kuaishou Inc.}
  \city{Beijing 100085}
  \country{China}}
\email{biaoli6@139.com}

\author{Peng Jiang}
\affiliation{%
  \institution{Kuaishou Inc.}
  \city{Beijing 100085}
  \country{China}}
\email{jp2006@139.com}

\def\authors{Jiayin Wang, Weizhi Ma, Chumeng Jiang, Min Zhang, Yuan Zhang, Biao Li, Peng Jiang}

\renewcommand{\shortauthors}{Jiayin Wang, et al.}


\begin{abstract}
In the era of information explosion, numerous items emerge every day, especially in feed scenarios. Due to the limited system display slots and user browsing attention, various recommendation systems are designed 
not only to satisfy users' personalized information needs but also to allocate items' exposure.
However, recent recommendation studies  mainly focus on modeling user preferences to present satisfying results and maximize user interactions, while paying little attention to developing item-side fair exposure mechanisms for rational information delivery. This may lead to serious resource allocation problems on the item side, such as the Snowball Effect. Furthermore, unfair exposure mechanisms may hurt recommendation performance.
In this paper, we call for a shift of attention from  modeling user preferences to developing fair exposure mechanisms for items.
We first conduct empirical analyses of feed scenarios to explore exposure problems between items with distinct uploaded times. This points out that unfair exposure caused by the time factor may be the major cause of the Snowball Effect.
Then, we propose to explicitly model item-level customized timeliness distribution, Global Residual Value (GRV), for fair resource allocation. This GRV module is introduced into recommendations with the designed 
Timeliness-aware Fair Recommendation Framework~(TaFR).
Extensive experiments on two datasets demonstrate that TaFR achieves consistent improvements with various backbone recommendation models. 
By modeling item-side customized Global Residual Value, we achieve a fairer distribution of resources and, at the same time, improve recommendation performance.

\end{abstract}

\ccsdesc[500]{Information systems~Recommender systems}

\keywords{Recommendation system, Item Fairness, Timeliness Distribution.}

\thanks{This work is supported by the Natural Science Foundation of China (Grant No. U21B2026, 62002191). Min Zhang is the corresponding author.}

\maketitle

\input{sections/S1-introduction}

\input{sections/S2-relatedWork}
\input{sections/S3-problem}

\input{sections/S4-model}

\input{sections/S5-application}

\input{sections/S6-conclusion}

\clearpage
\balance

\bibliographystyle{ACM-Reference-Format}
\balance
\bibliography{reference}

\end{document}

%% file: definitions.tex
\graphicspath{{figures/}}

%% file: packages.tex
\usepackage{booktabs} 
\usepackage{graphicx}
\usepackage{multirow}
\usepackage[inline]{enumitem}
\usepackage{hhline}
\usepackage{setspace}
\usepackage{makecell}
\usepackage{soul}
\usepackage{xcolor,colortbl}
\usepackage{threeparttable}
\usepackage{balance}
\usepackage{float}
\usepackage{threeparttable}
\usepackage[normalem]{ulem}
\usepackage{wrapfig}
\usepackage{arydshln}
\usepackage{algorithm}
\usepackage{algpseudocode}
\usepackage{amsmath}

\usepackage{graphicx}
\usepackage{subfigure} 

\usepackage{booktabs} 
\usepackage{diagbox}
 \usepackage{stfloats}
\usepackage{url}

\usepackage{hyperref}

%% file: sections/S1-introduction.tex
\section{Introduction}



In the age of information explosion, a large number of new items enter the candidate pool of recommendation systems each day, especially in feed scenarios~\cite{wu2022feedrec}. 
However, user browsing attention is limited and system display slots are scarce.
So various recommendation systems are designed to provide a personalized ranked list for each user from the growing candidate pool, which also determines items' exposure mechanisms, i.e. the allocation of resources.

An ideal recommendation system should not only meet users' distinct preferences for items, but also provide fair chances for the exposure of items.  
Most existing recommendation systems mainly focus on modeling users' preferences to generate satisfying recommendation results for users, which may provide more user-item interactions (e.g., clicks, durations, etc.) within the system. However, most of them pay inadequate attention to
item exposure mechanisms and ignore the item fairness issue.

\begin{figure}[htbp]
    \centering
    \subfigure[MIND News Dataset]{
	\includegraphics[width=0.46\linewidth]{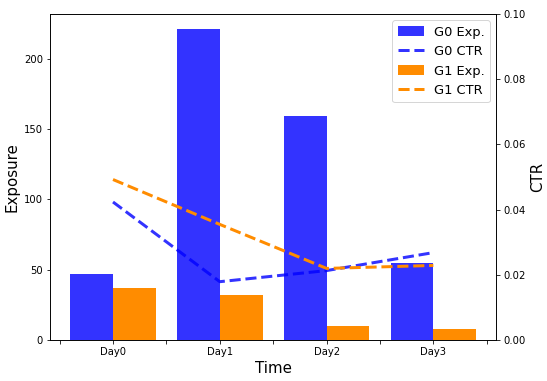}
        \label{fig:MIND_dynamic}
    }
    \hspace{0pt}
    \subfigure[Kuai Short Video Dataset]{
    \includegraphics[width=0.46\linewidth]{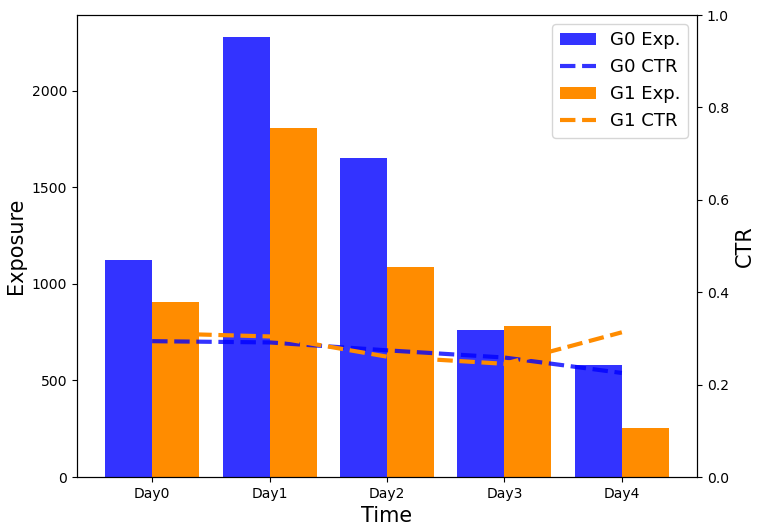}
        \label{fig:Kwai_dynamic}
    }
    \caption{Items uploaded earlier constantly receive more exposure opportunities: G0 contains entire items uploaded at 4 PM and G1 is all the items uploaded at 5 PM in two real-world systems.}
    \label{fig:intro}
\end{figure}

The lack of rational item exposure mechanisms could lead to severe problems in real scenarios.
For instance, early uploaded items are more likely to receive more resources, such as exposures, due to the Snowball Effect, which entails a self-amplifying process that originates from a minor impact and grows in magnitude over time, potentially resulting in either negative outcomes. In contrast, items uploaded recently often encounter the cold-start challenge and struggle to receive exposure opportunities compared to established items that currently dominate recommendation systems.
Figure~\ref{fig:intro} presents exposure distribution between two groups of items, which are uploaded in two consecutive hours. They have comparable user feedback (CTR), but the newly uploaded group (G1, in orange) receives significantly less exposure.
We dub this unfair exposure between items uploaded at different times as the \textit{time-sensitive unfair exposure} issues.
It will negatively affect the spread of time-sensitive information and further influence the quality of recommendations, which will eventually hurt both user satisfaction and the passion of item providers.

It is worth noting that our goal is to establish a fair competitive environment among items with varying upload times, rather than prioritizing the recommendation of a particular subset, such as newly uploaded items, or suppressing a certain portion of items, such as the current dominant items.
 Considering that each item undergoes a cold-start process,
 every item in the system will benefit if our approach can mitigate the Snowball Effect in the recommendation loops and allocate a fair amount of resources to fresh items.

The time-sensitive unfair exposure problem is a type of item fairness issue, which is different from another type of exposure-related study, i.e., popularity de-biasing.
The reason is that de-biasing work mostly tackles popularity bias for higher recommendation accuracy without concurrently considering item exposure fairness~\cite{zhang2021causal,wang2022make}. 
Existing studies on item fairness point out that for a fair recommendation system, each item should get resources proportionally to its utility. 
They have explored fair exposure mechanisms but treated item utility as a static score that ignores its global timeliness trends~\cite{steck2011item,10.1145/3397271.3401100,pitoura2021fairness,wang2022survey,li2022fairness}.
Thus they are unsuitable for handling this problem. 

In this study, we call for a shift of attention from user-side overly modeling to item-side fair exposure mechanism designs.
We conduct empirical analyses on two real-world feed datasets to show the unfair exposures between items with different upload times. Analyses show that the Snowball Effect can be mitigated by improving the competitiveness of newly uploaded items and removing excessive exposure for dominant items. At the same time, we observe that different items have their own trends of timeliness decay in feed scenarios.
Based on these findings, we then propose to explicitly model the item-level timeliness distribution, namely \textbf{Global Residual Value}, for fair resource competition between items uploaded at different times and design the \textbf{T}imeliness-\textbf{a}ware \textbf{F}air \textbf{R}ecommendation Framework (\textbf{TaFR}).
With abundant experiments, we show that more exposure resources can be fairly allocated to support the neglected new items, and recommendation quality is improved simultaneously.

To summarize, our main contributions are as follows:
\begin{enumerate}[leftmargin=*,nosep]
    \item We conduct empirical analyses on unfair exposure issues between items with different upload times in feed recommendation scenarios, which demonstrate that unfair exposure caused by the time factor may be the major cause of the Snowball Effect and items have diverse timeliness trends.
    \item We propose to explicitly model items' Global Residual Value, which is defined as the item-level customized timeliness distribution. The GRV module is designed to alleviate the time-sensitive unfair exposure issues (e.g., Snowball Effect) and provide a fair way for competition between items in exposures. The proposed GRV module is flexible to work with various backbone recommendation methods in our proposed Timeliness-aware Fair Recommendation Framework. 
    \item Extensive experiments with three types of backbone models demonstrate the effectiveness of TaFR in achieving both fairer exposure mechanisms and recommendation performance improvements.
\end{enumerate}


%% file: sections/S2-relatedWork.tex
\section{related Work}
\subsection{Fairness-aware Recommendation}
\label{sec:fairness}

Fairness in recommendation systems can be divided into three major categories by research subjects: user fairness, item fairness, and joint fairness (also called market fairness~\cite{fairness0})~\cite{pitoura2021fairness,wang2022survey,li2022fairness}. User fairness seeks equal treatment, in recommendation accuracy, explainability, etc., among different (groups of) users~\cite{ekstrand2018all,fu2020fairness}. Item fairness concerns whether the system treats items fairly~\cite{biega2018equity,rastegarpanah2019fighting}, such as equal prediction errors or resource allocations. They mainly treated item utility as static and have not modeled its utility changes over time, namely timeliness.
Work \cite{mansoury2020fairmatch} shows that
improvements in item fairness are possible to increase recommended diversity. \cite{feeds0} introduces context bias in feed recommendations for unbiased learning to rank. 
Note that fairness considerations are different with de-biasing for recommendation accuracy, as works focused on mitigating popularity bias without simultaneous considerations of fairness metrics might result in negative impacts on item fairness~\cite{zhang2021causal,wang2022fairness}.

In this paper, we focus on item-side time-sensitive exposure fairness issues, especially in feed systems where items have diverse timeliness trends. We address this unfair exposure problem, such as the Snowball Effect, by explicitly modeling items' customized timeliness across time in macroscopic views, instead of using static utility or absolute upload time for timeliness portrayal.


\subsection{Time-sensitive Recommendation}

Time-sensitive recommendation tasks aim to recommend the most desirable item at the right moment~\cite{du2015time}, in which temporal patterns are important to both user and item modelings.
In particular, feed recommender platforms (for news, weblogs, short-video, etc.), providing users with frequently updated content, are a typical type of time-sensitive systems ~\cite{okura2017embedding,wu2022feedrec}.
In time-sensitive recommendation tasks, time-aware algorithms are often used~\cite{gantner_factorization_2010,wang_make_2020,wang_time-aware_2021}. Survey~\cite{campos_time-aware_2014} evaluates the state of the art on time-aware recommendation systems (TARS), which deal with the time dimension in user modeling and recommendation methods. 
We incorporate the time-aware sequential model, TiSASRec~\cite{li2020time}, as one of the backbone models for comparisons. 
Note that most previous works are concentrated on users' browsing sequences and dynamic interesting modelings~\cite{agarwal2010fast,sun2016time,an2019neural,aliannejadi2019joint}, putting little attention on timeliness modeling on the item side.

This paper focuses on feed recommendation systems, which is a typical type of time-sensitive scenario with large-scale and frequently updated items. These features raise great challenges in the fair exposure allocation between items uploaded at different times. In this work, we aim to alleviate this time-sensitive unfair exposure with minimal negative or even positive impacts on recommendation accuracy.



\subsection{Item Timeliness Modeling}
\label{sec:item_modeling}
Item timeliness modeling is equally important with the more thoroughly researched user dynamic interests modeling in time-sensitive recommendation scenarios, as user and item together form an information system, and both have temporal characteristics (item's diverse timeliness features are introduced in empirical analysis in Section~\ref{sec:item_ana}).
\cite{wang2019modeling} models the dependencies between items inside a session and re-ranking works also focus on the pair-wise or list-wise item-item relationship constructions~\cite{pei2019personalized,abdollahpouri2019managing,pang2020setrank}.
Work~\cite{huang2022different} focuses on the age of an item and its effect on selection bias and user preferences for enhancements in rating predictions and work~\cite{wang2013opportunity} uses survival analysis in recommended opportunities modelings for accuracy enhancements in e-commerce scenarios.
Survival analysis, also called time-to-event analysis \cite{survival_analysis0,jenkins2005survival,wang2019machine}, is a branch of techniques that focused on lifetime modelings, such as the death in biological organisms, failure in mechanical systems, and user churns in games~\cite{li2021difficulty}.
Among all related techniques, the Cox proportional hazards model is a widely used procedure for modeling the relationship between covariates to survival or other censored outcome \cite{COX0,moolgavkar2018assessment,matsuo2019survival}.
In this paper, inspired by the survival analysis, we define items' Global Residual Value as the timeliness distribution and design the GRV modeling module for fair recommendations.

%% file: sections/S3-problem.tex
\section{Empirical Analyses in Feed Recommendation}
\label{Sec:analysis}

In order to better understand the time-sensitive unfair exposure issues in recommendation scenarios, we conduct empirical analyses on two real-world recommendation systems, which are introduced in Section~\ref{sec:two_dataset}.
In Section~\ref{sec:system_ana}, we first investigate the exposure situation between items uploaded at different times, and then evaluate the degree of exposure unfairness at the system level.
Section~\ref{sec:item_ana} presents diverse item-level timeliness trends, which demonstrate that different items in feed systems have their own timeliness distribution over time.
Based on these analyses, we call for further modeling of item-level customized timeliness distribution, defined as Global Residual Value, for fair exposure mechanisms.

\subsection{Two Feed Scenarios}
\label{sec:two_dataset}


Feed recommendation systems are typical time-sensitive scenarios, where items evolve rapidly, and delivering timely content is a system requirement.
We select a PC-side news recommendation website and a mobile-side short video social application as representative scenarios for analysis and experiments in this paper:

\begin{itemize}[itemsep=1pt,parsep=0.8pt]
  \item \textbf{MIND}~\cite{wu2020mind}: the publicly available dataset collected from anonymized behavior logs of the \href{https://microsoftnews.msn.com/}{Microsoft News} website. 
  \item \textbf{Kuai}: desensitized impression log of a mobile application for short-video recommendation and the collected data, with considerable size and diverse types of real users' feedback, will be publicly released along with this paper after acceptance.
\end{itemize}

\subsection{Analysis on Items' Exposure Distribution}
\label{sec:system_ana}
There are numerous items in the candidate pools of recommendation systems and they are uploaded at different times. Items uploaded at earlier times may accumulate more exposure and user feedback, leading to precise modeling and becoming dominant in the system. Newly uploaded items are experiencing cold start problems, mainly resulting in tentative recommendations. In feed recommendation scenarios, it is critical to deliver updated information and iterate over outdated items.

In this work, we examine the competition for exposure resources between items uploaded at different times on the two real-world platforms.
Exposure opportunities are adopted as resource measurements since they are the most important system-controlled benefits for items. For item utility (also called merit)~\cite{biega2018equity,10.1145/3397271.3401100}, CTR is adopted as the user feedback metric. In the following part, we evaluate the time-sensitive unfair exposure issues through an item-level case study and system-level assessment.

\subsubsection{\textbf{Item-level Case Study}}
\ 
\newline
Figure \ref{fig:intro} in the Introduction illustrates the exposure and user feedback situation between two groups of items uploaded in two consecutive hours. Group 0, colored in blue, contains all the items uploaded in 4PM and items uploaded in 5PM belongs to Group 1, which is in orange. 
Dashed lines reflect the user feedback and solid bars show the system-allocated exposures in the following days. Although Group 1 has comparable or even higher CTR, it received less exposure compared with the early uploaded and currently prevailing group in Day 1. The suppression trend continue to hold in the following days, and eventually, both group lose timeliness.



This case study shows that in real-world recommendation systems, there are situations where established item crowd out resources of new items. This trait is not conducive to items' cold start and the dissemination of fresh content, especially in feed systems. It can also mix with the Snowball Effect, creating a pernicious circle.
In the following analyses, we look further at the prevalence and severity of this time-sensitive unfair exposure allocation problem from a holistic perspective.

\subsubsection{\textbf{System-level Evaluation}}
\ 
\newline
To gain further insight into the system-wide picture, we group all items according to their upload time and evaluate the subsequent performance of each group in recommendation systems.
\begin{figure}[htbp]
    \centering
    \subfigure[MIND news dataset]{
	\includegraphics[width=0.472\linewidth]{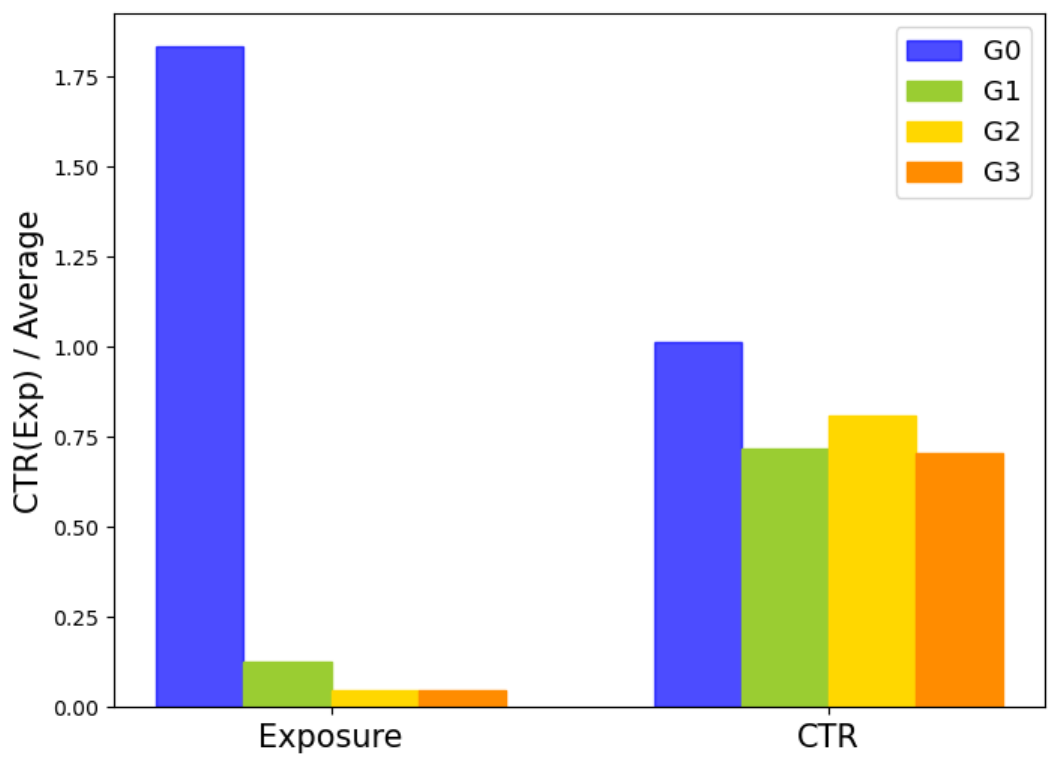}
        \label{fig:MIND_group}
    }
    \hspace{0pt}
    \subfigure[Kuai short video dataset]{
    \includegraphics[width=0.472\linewidth]{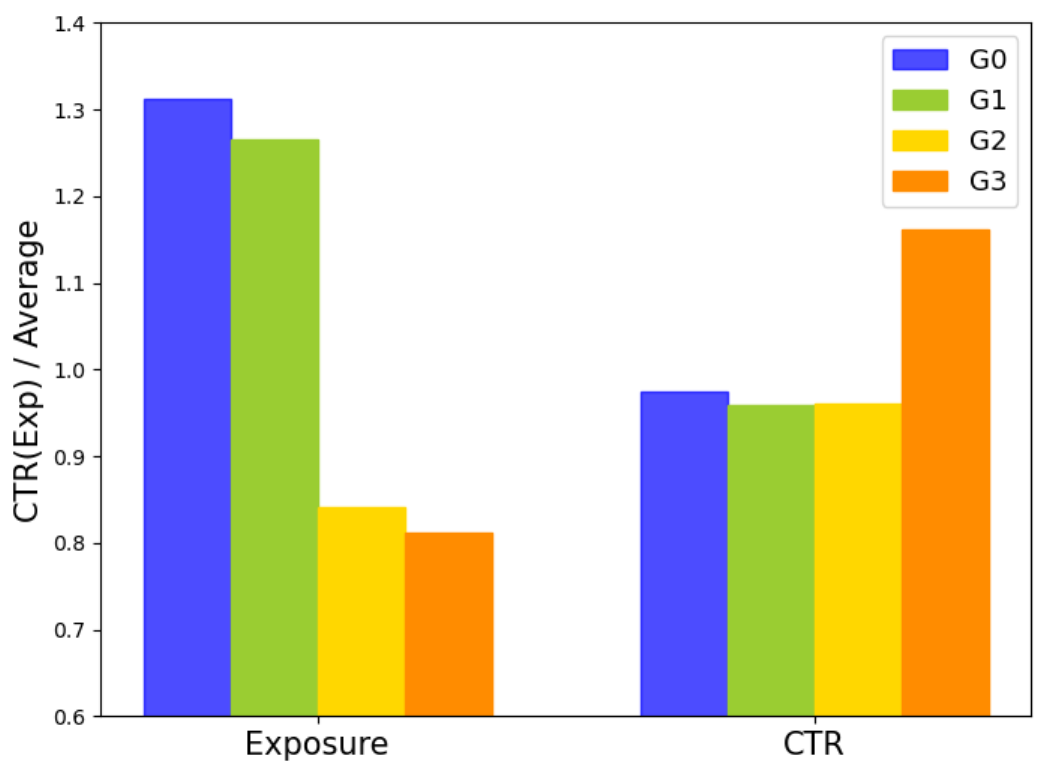}
        \label{fig:Kwai_group}
    }
    \caption{Time-sensitive unfair exposure: G0-G3 contain all items uploaded on four consecutive equal lengths of time, respectively.
    As upload times move backward (larger group number), exposure opportunities decrease.}
    \label{fig:analysis_two}
\end{figure}
Results are shown in Figure \ref{fig:analysis_two}. As the group number increases, the items belonging to the group are uploaded more recently. 
Specifically, for MIND, we group items first exposed on Day 1 12 AM to 12 PM into 4 groups, and each group contains items uploaded in 3 hours. We evaluate each group's performance from Day 1 12 PM to Day 2 12 AM. For Kuai, we group items uploaded on Day 1 into 4 groups with each group uploaded within 6 hours, and examine their performance on the following Day 2.
Each group's CTR and exposure situations in the following observation period are plotted. Both metrics are normalized as follows: 
\begin{equation}
Y_{Exp}=\frac{\overline{Exp}_{i\in G_x}}{\overline{Exp}_{i \in G}}, ~
Y_{CTR}=\frac{CTR_{i\in G_x}}{CTR_{i \in G}}
\end{equation}
First we calculate the average exposure (or CTR) of items in Group x, and we divided it by the average exposure (or CTR) of the system.
For MIND, the earliest uploaded groups crowd the most exposure opportunities in the system while each group has comparable user feedback.
The same time-sensitive unfair exposure issue holds still for the Kuai dataset, where the latest uploaded item group has the highest user feedback while receiving the least resources.

As shown above, unfair exposure allocation issues are prevailing in recommendation systems, especially in scenarios with frequently uploaded new items. The unfairness might build up like snowballs, resulting in serious information churn and further affecting user satisfaction.
In the next section, we further dive into item-level perspectives for causes and mitigation methods of these time-sensitive unfair exposure problems.

\subsection{Diverse Timeliness Characteristics of Items}
\label{sec:item_ana}

Analyses in the previous subsection identify the time factor as one of the causes and solutions of unfair exposure problems, such as the Snowball Effect.
Focusing on this idea, we further investigate the major item-level time factor in feed scenario, timeliness, which is related to while different from items' uploaded time.

\begin{figure}[htbp]
 \centering 
    \subfigure[Item A]{
    \includegraphics[width=0.47\linewidth]{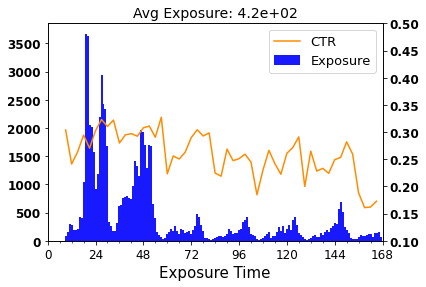}
        \label{fig:0-0}
    }
    \subfigure[Item B]{
	\includegraphics[width=0.47\linewidth]{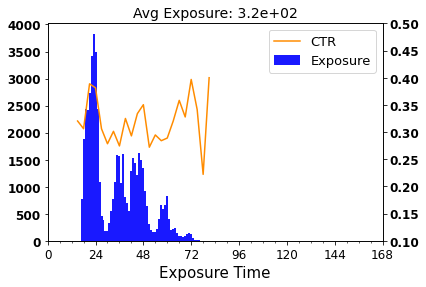}
        \label{fig:0-1}
    }
    \caption{Items with diverse timeliness patterns in Kuai dataset.}
    \label{unfair:case_dynamic}

\end{figure}

For further understanding of the item-level timeliness characteristics, we carry out the case study in feed scenarios.
Figure~\ref{unfair:case_dynamic} gives out two typical items' performances and system-allocated exposure resources across a time range of 7*24 hours in Kuai dataset.
Item performances are plotted in orange and bars in blue present the system-allocated exposure amounts.
Note that the scales of the exposure coordinate axes are inconsistent and we provided the average exposure amounts per hour on the top. 

\subsubsection{\textbf{Time-sensitive Unfair Exposure}} 
\ 
\newline
Comparing Figure~\ref{fig:0-0} and~\ref{fig:0-1}, which are uploaded on the same day while getting their first exposures in less than 12 hours difference, the former gets more exposures in the early stages and keeps getting system exposures even if it has relatively low user feedback in the following days. 
These cases further verify the presence of the Snowball Effect. If an item accumulates a considerable amount of resources at early time stages, it tends to enjoy greater privileges in the subsequent competition.

\subsubsection{\textbf{Item-level Diverse Time Patterns}}
\ 
\newline
In order to mitigate unfair exposure situations, such as the Snowball Effect shown above, recommendation systems need to support newly uploaded items and remove excessive exposure to currently dominant items.
We need to model the timeliness of items in a macroscopic view. This is different from items' uploaded time or exposure amounts since different items have different time patterns. Some headline items might have strong timeliness in feed scenarios, quickly replaced by new contents, and some might be time-insensitive, possessing long spread periods. Figure~\ref{unfair:case_dynamic} gives the examples of these two kinds of items in our Kuai dataset.

Based on the above analyses of item-side exposure situations and items' diverse timeliness characteristics, we propose to model an item-level customized dynamic descending value (namely \textit{Global Residual Value}) to model their timeliness distribution at each time point.
In the next section, we introduce the definition and modeling module of the proposed \textit{Global Residual Value}.

%% file: sections/S4-model.tex
\section{Global Residual Value Module}
\subsection{Task Statement}

Given a newly uploaded item $i$ and its interaction history in the observation period of duration $T_{obs}$,
the item dynamic modeling task is to forecast $i$'s changing global timeliness distribution, $GRV_i(t)$, which is the probability of keeping active (i.e., having of timeliness or interactions) at each future moment. This value is different from the user-item level relevance score as GRV portrays the timeliness of content at the system level. Table \ref{tab:notation} displays the main concepts used in the Global Residual Value module.


\begin{table}[htbp]
\caption{Notations used in the Global Residual Module.}
    \centering
    \begin{tabular}{c|c}
    \hline
    Notations & Explanations \\ \hline
    $T_{obs}$ & duration of the observation period \\
        $T_{pred}$ & duration of the prediction period \\
    $T_{i,0}$ & upload time of item $i$\\
        $T_{i,d}$ & deactivate label time of item $i$ \\
    $\mathbf{F_i}$ & interaction history in observation period \\
    $S_i(t)$ & timeliness status of item $i$ at time $t$\\
    $h_i(t)$ & deactivate hazard by survival analysis at time t \\   
    \hline
    \end{tabular}
    \label{tab:notation}
\end{table}

To calculate items' GRV, we need to utilize item interactions in the systems, so we define the mentioned observation period to collect interactions of each item from upload time $T_{i,0}$ to $T_{i,0}+T_{obs}$. $T_{obs}$ is a hyper-parameter in distinct scenarios. Then, we can use the collected information to forecast the GRV of the prediction period ($T_{i,0}+T_{obs}$ to $T_{i,0}+T_{obs}+T_{pred}$).


 \subsection{GRV Modeling}
 \label{sec:GRV_define}
First, we formally define the Global Residual Value of item $i$ at future time point $t$ (t $\geq$ 0), which is the quantitative measurement of timeliness:

\begin{equation}
\label{equ:GRV_definition}
\begin{aligned}
GRV_i(t)=P(S_i(t)=\text{active}) = P(T_{i,d}>t|\mathbf{F_i})
  \end{aligned}
\end{equation}
where $S_i(t) \in \{\text{active}, \text{deactivated}\}$, 
$\mathbf{F_i}=\mathbf{F_i}(\mathbf{t}) \in \mathrm{R}^{T_{obs}*|F|}$, $T_{i,d} \in \mathrm{R}$. 
The Global Residual Value measures the probability that an item will remain active at a time point, which is equivalent to the deactivation event occurring after the time point $t$.
By capturing item-level timeliness evolution patterns in the system, we use items' performances in the observation period to predict the future Global Residual Value. 
Note that $T_{i,d}$, the deactivation time of items, can be manually labeled by administrators or automatically generated by systems. Subsection \ref{Sec:deactivation} introduces a design for the system determination method used in this paper, with no additional labeling requirements.

\begin{figure*}[htbp]
    \centering
\includegraphics[width=0.95\linewidth]{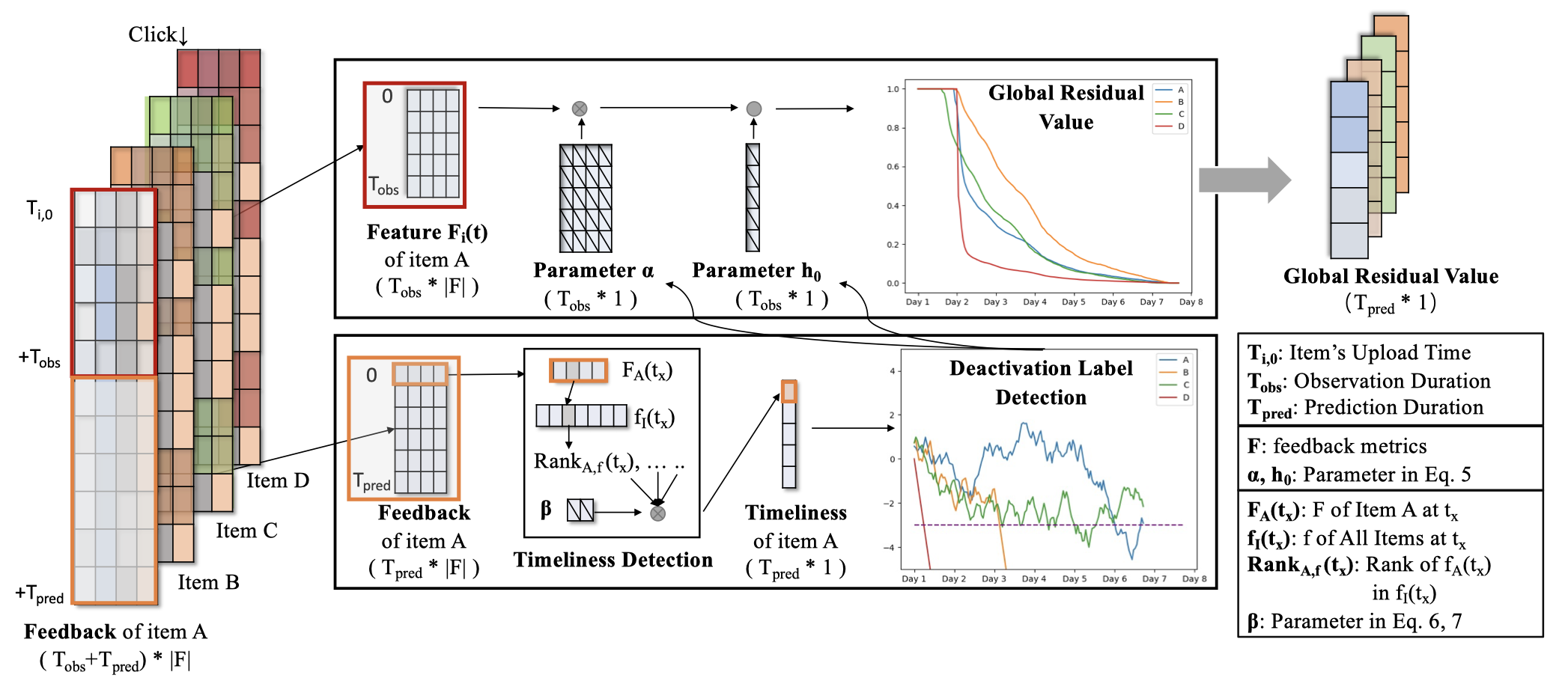}
\caption{The Global Residual Value Module, which models the item's timeliness distribution in the prediction time period with its past performances in the observation window. The input is item $i$'s user feedback from upload time $T_{i,0}$ to the end of observation period $T_{i,0}+T_{obs}$. The output is Global Residual Value vector $\mathbf{GRV_i}$ with the length of $T_{pred}$.}
    \label{fig:model}
\end{figure*}

As the Global Residual Value can not be directly calculated, inspired by survival analysis, we further construct its relationship with the 
hazard probability, $h_i(t)$, which is defined as the following:
\begin{equation}
\label{equ:hazard_define}
\begin{aligned}
h_i(t)=\lim _{\delta(t) \rightarrow 0} \frac{\operatorname{P}(t \leq T_{i,d} < t+\delta(t) \mid T_{i,d} > t)}{\delta(t)}
\end{aligned}
\end{equation}

Equation~\ref{equ:relationship} portrays the relationship between hazard function $h_i(t)$, accumulated hazard function $H_i(t)$ and our designed $GRV_i(t)$. As shown below, by modeling the hazard function, we are able to obtain the Global Residual Value. 
\begin{equation}
\label{equ:relationship}
\begin{aligned}
H_i(t) =\int_0^t h_i(x) dx &=\int_0^t -\frac{GRV'_i(x)}{GRV_i(x)} dx =-\log (GRV_i(t))\\
GRV_i(t)&=\exp (-H(t))
\end{aligned}
\end{equation}

To further model the hazard function,
we introduce the assumption in Cox's proportional hazard model~\cite{jenkins2005survival}, which is that the log hazard of an individual is a linear combination of its covariates and a system-level time-varying benchmark hazard.
According to \cite{stensrud2020test}, we can adopt this assumption in our item timeliness modeling task and construct $\mathbf{h_i}$ as follows:
\begin{equation}
\label{equ:GRV_calculation}
\begin{aligned}
h_i(t)=h_0(t)*Exp[\sum_{t_{x}=T_{i,0}}^{T_{i,0}+T_{obs}}\sum_{f \in F} \alpha_f(t_{x})*(f_i(t_x)-\overline{f_{I}(t_{x})})]\\
t \in (T_{i,0}+T_{obs},T_{i,0}+T_{end}]
\end{aligned}
\end{equation}
where we use the item's performance in the observation window, $\mathbf{F_i}$, as covariates.




 \subsection{Module Overview}
The Global Residual Value module is described in Figure~\ref{fig:model}. 
The following parts illustrate the inference and training procedures of our designed Global Residual Value modeling module.

\subsubsection{\textbf{Inference}}
\
\newline
The upper box in Figure~\ref{fig:model} is a diagram of the inference process. The input is items' past performance in the system, marked red on the upper left in the figure, and the module predicts the item-level customized Global Residual Value vector across time. More specifically, when an item is uploaded at time $T_{i,0}$ and passes its observation period (from $T_{i,0}$ to $T_{i,0}+T_{obs}$), its performance in the system is transport to the GRV module as its features for timeliness distribution calculations. Items' performance, $\textbf{F}$, contains various user feedback such as CTR, watch ratio, like rate, etc.
The module follows Equation~\ref{equ:GRV_calculation} to forecast $i$'s Global Residual Value vector in the prediction period~(from time $T_{i,0}+T_{obs}$ to $T_{i,0}+T_{obs}+T_{pred}$). This vector is further sent into Timeliness-aware Fair Recommendation Framework introduced in Section~\ref{sec:recFramework}.

\subsubsection{\textbf{Training}}
\
\newline
The training process is illustrated in the lower box in Figure~\ref{fig:model}. It contains mainly two parts: the deactivation label generation and the parameter learning.

GRV module collects all the impression data from items' upload to $T_{i,0}+T_{obs}+T_{pred}$, where the subsequent performance after the observation period, $[T_{i,0}+T_{obs}$,$T_{i,0}+T_{obs}+T_{pred})$, marked orange in the lower left of the framework figure, is unreachable at the inference time point. Logs in this period are sent to the timeliness detection part. Note that items' deactivation of timeliness can be manually labeled by administrators in feeds systems or automatically determined by strategies. Subsection~\ref{Sec:deactivation} introduces the method we adopt for deactivation label detection. 

If the deactivation of an item does not trigger during the whole observation and prediction period, we cannot get the deactivation label time and this item forms a censored case. Since censors might occur, it is not appropriate to use the loss function such as mean squared error or mean absolute error loss for GRV module learning. We maximize the log-likelihood in module training to get the parameter $\mathbf{h_0}$ and $\boldsymbol{\alpha}$ in Equation~\ref{equ:GRV_calculation}.

The following subsection introduces the deactivation label detection method adopted in our module. This label ($T_{i,d}$) is used in the training process. Note that it can also be labeled manually according to the system's needs for content timeliness.







\subsubsection{\textbf{Deactivation Label}}
\label{Sec:deactivation}
\ 
\newline
The deactivation alert can be triggered up to once for each item. The occurrence of an alert means that the item's global timeliness is close to 0 and should be eliminated from the time-sensitive system.
The deactivation trigger time is marked as $T_{i,d}$.
If the item does not trigger any alerts within the whole time period of $[T_{i,0}, T_{i,0}+T_{obs}+T_{pred})$, it will become a censored case and will be removed from modeling training.

\begin{equation}
\begin{aligned}
v_i(t)=\mathbf{1}_{Exp_i(t)>0}*(R_i(t)-\beta_E) + \mathbf{1}_{Exp_i(t)=0}* (-\beta_{nE})
\\
R_i(t)=Rank(f_i(t),f_I(t))
\label{equ:vital}
\end{aligned}
\end{equation}
\begin{equation}
\begin{aligned}
T_{i,d}= \mathop{\arg\min}\limits_{t} \, \sum_{t_x=T_{i,0}}^t v_i(t_x) < \beta_d
\label{equ:alert_time}
\end{aligned}
\end{equation}
The deactivation label generation mechanism is designed under item-level feedback. We give each item a vitality score under Equation \ref{equ:vital}. $\mathbf{1}_{f}$ is the indicator function, $Exp_i(t)$ is the exposure amounts of item $i$ at time period t, $R_i(t)$ is the percentile rank of i's feedback in the system at time t, $\beta_E$ and $\beta_{nE}$ are two hyper-parameters. As we gain the item's evaluation at each time granularity, which ranges from min($-\beta_E$, $-\beta_{nE}$) to $1-\beta_E$, we add up this score to get the cumulative vitality performance. If it is less than the threshold $\beta_d$, then the item triggers the deactivation alert at this time point. This module is presented in the bottom part in Figure~\ref{fig:model}.
We need to control the censoring rate by appropriately designing the deactivation detection mechanism and selecting appropriate hyper-parameters to replace manual labels. Detailed parameter settings are tested and introduced in experimental settings in Section~\ref{sec:exp_settings}.


\section{Timeliness-Aware Fair Recommendation}
\label{sec:recFramework}
In this section, we introduce the calculated GRV of items into the top-K recommendation models for fair exposure allocations. 
We use $r_x=(u_x, t_x)$ to denote the request by user $u_x$ in time $t_x$, where each request $r_x$ consists of a user id and inquiry time. Recommendation systems receive a request list, $R=\{r_1, r_1, ..., r_{n}\}$, and respond to each request, producing the recommendation list $O=\{L_1, ... L_n\}$ with $L_x=[i_{x,1},..., i_{x,m}]$. These lists are consumed (i.e., read or watched) by users in time sequences, and the corresponding user feedback is recorded by systems for further modeling.
Viewed from the global system level, items, with different exposed hours and timeliness trends, get resources in the order of $[i_{u_1,pos_1},..., i_{u_n,pos_m}]$ with a length of $|R|*|L|=n*m$. 





Given a user request $r_x=(u_x,t_x)$ and candidate item set $I$, 
TaFR generates the recommendation list based on the pre-calculated Global Residual Value, $\mathbf{GRV_I}$, and the backbone model produced recommendation scores between items $I$ and the user $u_x$ for personalized and fair exposure.

\begin{figure}[htbp]
    \centering
    \includegraphics[width=0.95\linewidth]{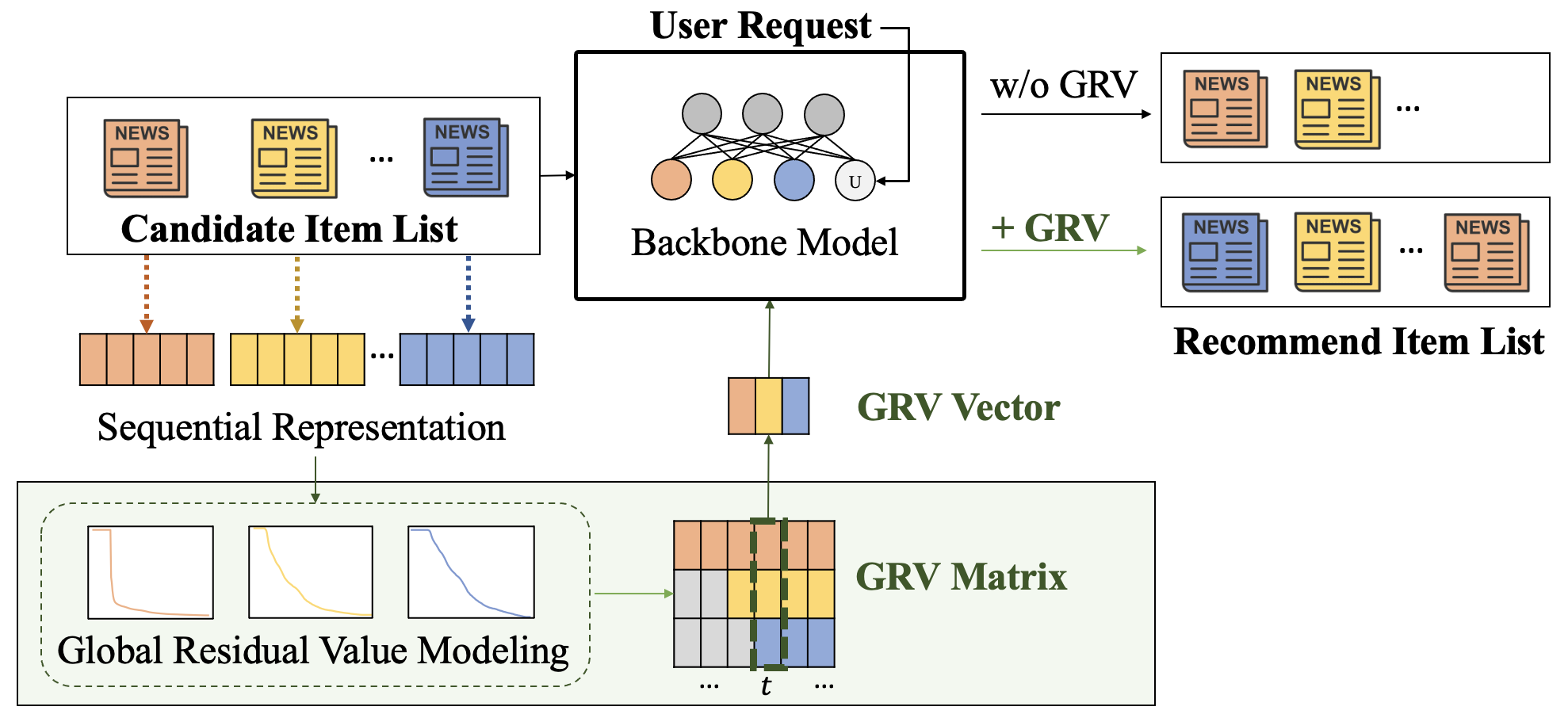}
    \caption{Timeliness-aware Fair Recommendation Framework explicitly models items' Global Residual Values for personalized recommendations and fair exposure allocations.}
    \label{fig:framework_overview}
\end{figure}

\subsection{Framework Overview}
\label{sec:frameworkOverview}


We design the \textbf{T}imliness-\textbf{a}ware \textbf{F}air \textbf{R}ecommendation Framework, \textbf{TaFR}\footnote{https://github.com/Alice1998/TaFR}, aiming to alleviate the time-sensitive unfair exposure in recommendation scenarios with minimal negative or even positive impacts on recommendation accuracy.
The overall structure of TaFR is shown in Figure \ref{fig:framework_overview}. It is composed of two parts: the backbone model for personalized recommendations and the Global Residual Value Module for items' timeliness distribution modeling.

Typical recommendation systems generate output (exposure list $L_x$) based on the user request $r_x=(u_x, t_x)$ and the candidate item set $I$. In this process, each item is assigned a certain recommendation score (mainly representing the degree of relevance between the user and item at time t). 
When allocating exposure opportunities, each exposure list is generated individually based on the predicted recommendation scores for the current user request.
The system-level deviations of recommendation scores, especially unfair competitions between items with different upload times and timeliness patterns,
are not properly and explicitly considered.

In TaFR, we utilize the personalized recommendation models as backbone modules and integrated them with the item-level customized timeliness distribution pre-calculated in the Global Residual Value module. These two modules and their integration methods are introduced in the following subsections. Table~\ref{tab:framework_notion} displays notations used in TaFR.

\begin{table}[htbp]
\caption{Notations used in the framework.}
    \centering
    \begin{tabular}{c|c}
    \hline
    Notations & Explanations \\ \hline
    $BBM_{r(u,t)}(i)$    & Backbone module \\
    $\mathcal{B}(u,i,t)$    & Backbone modeling algorithms \\
    $GRV_I(t)$    & Global Residual Value module \\
    $\mathcal{G}_i(t)$    & GRV modeling algorithms \\
    $P_i(t)$    & system predicted score for item $i$ at time $t$ \\
    $F_i(t)$    & users feedback for item $i$ at time $t$ \\ 
     $\gamma$    & hyper-parameter for GRV integration \\ 
    \hline
    \end{tabular}
    \label{tab:framework_notion}
\end{table}

\subsubsection{Backbone Recommendation Score}

\begin{equation}
\begin{aligned}
  BBM_{r(u,t)}(I)=\mathcal{B}(u(t),I,t)
  \end{aligned}
\end{equation}

\looseness=-1 
Backbone module $BBM_{r(u,t)}(I)$ produces recommendation scores based on the user request $r(u,t)$ and item candidate set $I$. The backbone algorithm $\mathcal{B}$,
requires user-item interaction logs for the id-based recommendation score prediction. Note that TaFR has no special requirements for backbone models and most recommendation algorithms can be used as the backbone. $BBM_{r(u,t)}(I)$ module mainly focuses on improving recommendation accuracy and lacks macroscopic consideration of fair item exposure mechanisms.

\subsubsection{Global Residual Value}
\ 
\newline
The Global Residual Value is defined in Section~\ref{sec:GRV_define} as items' global timeliness distribution. It represents the probability that potentially interested users for item $i$ exist among the unexposed users at time $t$. This could be quantified as the probability that the item is active at the current $t$ and is modeled by the $\mathcal{G}_i$ as shown in Equation \ref{equ:RRV_define}.
\begin{equation}
\label{equ:RRV_define}
\begin{aligned}
GRV_i(t)=P(S_i(t)=\text{active})=\mathcal{G}_i(t-T_{i,0})
  \end{aligned}
\end{equation}

In the feed recommendation scenario, items typically obtain certain timeliness patterns as illustrated in Section \ref{Sec:analysis}. More specifically, at the end of observation time $T_{i,0}+T_{obs}$, the system acquires user feedback on the item $i$ from the time it is uploaded ($T_{i,0}$) to the current moment. Usually, the system also predicts items' relevant scores with candidate users. These two types of information, predicted recommendation score $\textbf{P}$ if available, and real user feedback $\textbf{F}$, are used as the input factors in Global Residual Value Module. This modeling process $\mathcal{G}$ is shown in Equation \ref{equ:RRV_model}.

\begin{equation}
\label{equ:RRV_model}
  \begin{aligned}
GRV_i(t) &=\mathcal{G}\big(P_i(T_{i,0}), F_i(T_{i,0}); \cdots ; \\ &P_i(T_{i,0}+T_{obs}-1), F_i(T_{i,0}+T_{obs}-1); P_i(T_{i,0}+T_{obs})\big)(t) \\
&t \in \left[T_{i,0}+T_{obs},T_{i,0}+T_{obs}+T_{pred}\right]
\end{aligned}
\end{equation}

\subsection{GRV-based Recommendation}
The framework calculates personalized recommendation scores $BBM_{r(u,t)}(I)$ and Global Residual Value $GRV_I(t)$ for item timeliness modelings. TaFR combines the ratings from two perspectives for final rankings. In this work, the aggregation method is shown in Equation~\ref{equ:combination}. 

\begin{equation}
\label{equ:combination}
  \begin{aligned}
\mathcal{G}_{r(u,t)}(I) &=(1-\gamma)*BBM_{r(u,t)}(I) + \gamma* GRV_i(t) \\
&=(1-\gamma)*\mathcal{B}(u(t),I,t) + 1-\gamma* \mathcal{G}_I(P_i(\mathbf{t_P});F_i(\mathbf{t_F}))(t-T_{I,0}) \\
& t \in [T_{I,0}+T_{obs},T_{I,0}+T_{obs}+T_{pred}]\\
&\mathbf{t_P} = [T_{I,0}:T_{I,0}+T_{obs}], ~\mathbf{t_F} = [T_{I,0}:T_{I,0}+T_{obs})
\end{aligned}
\end{equation}
$\gamma$ is the weight between user preferences and item  timeliness. $T_{obs}$ is the hyper-parameter for the length of the observation window and $T_{pred}$ is the duration for the prediction period. GRV module obtains items' user feedback and system scoring and at the end of the observation period forecasts items' timeliness distribution vector, $GRV_i(t)$, with the length of $T_{pred}$.

%% file: sections/S5-application.tex
\section{Experiments and Evaluations}
\label{sec:experiments}

As introduced in Section~\ref{sec:frameworkOverview}, TaFR can easily accommodate most personalized recommendation algorithms as backbones and is applicable to various time-sensitive recommendation scenarios.
In this section, we conduct extensive experiments with three types of backbone models on two datasets to verify the effectiveness of our framework in terms of recommendation accuracy and time-sensitive exposure fairness. Experimental settings are described in Section~\ref{sec:exp_settings}. Overall performances of TaFR are presented in Section~\ref{sec:exp_overallPf} and analysis of exposure fairness is shown in Section~\ref{sec:exp_fairness_analysis}. In Section~\ref{sec:exp_GRV}, we further examine the item GRV module to validate its ability on item-level customized timeliness modeling. 


\subsection{Experimental Settings}
\label{sec:exp_settings}

\subsubsection{\textbf{Datasets}}
\ 
\newline
We use two datasets described in Section~\ref{sec:two_dataset}.
The dataset settings for GRV module and recommendation framework are introduced below:

\begin{itemize}[itemsep=1pt,parsep=0.8pt]
  \item MIND~\cite{wu2020mind}: 
  We use the train and validation set in MIND, including 7 days' logs, and conduct a 10-core filter. For GRV, we randomly sample 20\% for testing and conduct recommendations on these 20\% items to prevent information leakage.
  \item Kuai:
  We collect and use items uploaded in two days and their exposure feedback in the following 7 days. For GRV, we use logs of items uploaded on the first day and use the second-day-uploaded items to compose the recommendation sets.
\end{itemize}

For GRV module, we use CTR as the only user feedback $\mathbf{F_i}$ and set the system predicted relevance score $\mathbf{P_i}$ to null for offline simulations due to dataset limitations.
For recommendations, on both datasets, we use the first 3 days for training and randomly split the last 2 days for validation and testing.
To better test TaFR's abilities for fair resource allocations, we conduct experiments based on the test-all settings. Note that this will lead to an overall decrease in the accuracy of the recommendation.

\subsubsection{\textbf{Framework Backbones}}
\ 
\newline
\looseness=-1 
TaFR has no special requirements for backbone models because the GRV module could integrate with most recommendation algorithms. We test three types of models as our backbones:
\begin{itemize}[itemsep=1pt,parsep=0.8pt]
  \item NeuMF~\cite{he2017neural}: neural collaborative filtering method.
  \item GRU4Rec~\cite{hidasi2015session}: sequential recommendation algorithm.
   \item TiSASRec~\cite{li2020time}: sequential and time interval aware recommendation.
\end{itemize}

\looseness=-1 
For each backbone algorithm, we test it only and integrate it with the normalized uploaded time and our Global Residual Value vector. Specifically, the upload time baseline (+time) replaces the GRV with the normalized upload time, $\mathbf{t_{i,exp}}$, which is calculated as follows:
\begin{equation}
\label{equ:time}
t_{i,exp}(t)=
1-\frac{t-T_{i,0}}{t - \text{min}~T_{I,0}+ \delta}=
\frac{T_{i,0}-\text{min}~T_{i,0}+\delta}{t - \text{min}~T_{I,0}+ \delta}
\end{equation}
Equation~\ref{equ:time} first normalizes $i$'s exposed time with the maximum item expose time at the current system and reverses it since longer expose time represents lower timeliness.

\subsubsection{\textbf{Parameter Settings}}
\ 
\newline
For the Global Residual Value module, we set the observation and prediction duration $T_{obs}+T_{pred}$ as 7*24 hours, which is one complete week, and the observation window $T_{obs}$ as 12 hours in MIND and 24 hours in Kuai due to different update frequency in news and short-video platforms. For hyper-parameters in GRV module~(Equation~\ref{equ:vital} and \ref{equ:alert_time}), aggregation settings in TaFR (Equation~\ref{equ:combination}) and the baseline method (Equation~\ref{equ:time}), we conduct pilot experiments for grid search in the range of [0, 0.5] with the step size of 0.1 respectively, $\beta_E$, $\beta_{nE}$ are set to 0.5 and the $\beta_d=-3$. $\gamma$ is set as 0.3, 0.3, 0.1 in MIND and 0.2, 0.1, 0.1 in Kuai for three backbones respectively. $\delta$ is set to 0, when $\text{min}~T_{I,0}=0$ we set the value  of $t_{i,exp}(t)$ as 0 in Equation~\ref{equ:time}.

\begin{table*}[htbp]
\caption{Overall Recommendation Results: The Timeliness-aware Fair Recommendation Framework (TaFR) is tested with three types of backbone models (NeuMF, GRU4Rec, TiSASRec) on two datasets (MIND and Kuai). For each backbone model, we test its performance alone and integrated it with upload time~($\text{TaFR}_{time}$) and our proposed Global Residual Value module~($\text{TaFR}_{GRV}$). This table presents the evaluation on recommendation accuracy, measured by Hit Rate (HR@k) and Normalize Discounted Cumulative Gain (NDCG@k), time-sensitive exposure fairness, evaluated by (New) Item Coverage ((N\_)Cov@k). Higher values of all metrics represent better results. ** represents p<0.05 significance and * means p<0.1 significance comparing with backbone model. For each dataset, the best performance row is marked in bold and the second best is underlined.}
\label{table:recommendation}
\centering
\resizebox{0.97\linewidth}{!}{
\begin{tabular}{cllrrrrrrrr}
\toprule
Dataset& BackBone & Method & HR@5  & NDCG@5 & N\_Cov@5 & Cov@5 & HR@10  & NDCG@10 &  N\_Cov@10 & 
 Cov@10  \\
\cmidrule(lr){4-5}  \cmidrule(lr){6-7}  \cmidrule(lr){8-9}  \cmidrule(lr){10-11}  
  \multirow{9}*{MIND} & \multirow{3}*{NeuMF} & backbone & 0.0953 & 0.0754 &  0.0026 & 0.0059& 0.1251 & 0.0846  &	0.0026 & 0.0099\\
   & & $\text{TaFR}_{time}$ & 0.0972  &	0.0762 & 0.0026 &0.0058	&0.1266  & 0.0855  &	0.0026 & 0.0087 \\
 & & $\text{TaFR}_{GRV}$&**\textbf{0.2204} &	**\textbf{0.1511} &  **\textbf{0.2992} & **\textbf{0.0657}	& **\underline{0.2403}	& **\underline{0.1574}  &	**\textbf{0.4698}  & **\textbf{0.1037}\\
\cmidrule(lr){4-7} \cmidrule(lr){8-11}
 & \multirow{3}*{GRU4Rec} & backbone & 
 0.1344&	0.0906 &0.0031&	0.0099 & 0.1626	&0.0995 &	0.0037 &\underline{0.0153} \\
  & & $\text{TaFR}_{time}$& 0.1302&	0.0909 & 0.0026 & **0.0067  &	0.1533 & 0.0982  &	0.0026& **0.0101  \\
& & $\text{TaFR}_{GRV}$& **\underline{0.2141}&	**\underline{0.1494}  & **0.0110& 0.0074 	&**\textbf{0.2453} &**\textbf{0.1592} &	**0.0147& 0.0113 \\
\cmidrule(lr){4-7} \cmidrule(lr){8-11}
  & \multirow{3}*{TiSASRec} & backbone & 
  0.1321&	0.0897 & 0.0698&\underline{0.0562} 	&0.1867	&0.1071 &	0.1050 & 0.0860 \\
   & & $\text{TaFR}_{time}$& 0.1239&	0.0808 & **0.0100 & **0.0134 	&0.1479&	0.0883  &	**0.0142 & **0.0216 \\
& & $\text{TaFR}_{GRV}$& **0.1764&	**0.1158& **\underline{0.1417} &0.0492  &	**0.2219&	**0.1303 & 	**\underline{0.2247}& 0.0779 \\ \midrule
 \multirow{9}*{Kuai} & \multirow{3}*{NeuMF} 
&backbone& 
0.2422	&0.1911	& 0.1128 & 0.6309 &0.3114&	0.2134 & 0.1579 &0.7013  \\
& &$\text{TaFR}_{time}$& **0.2255 & **0.1641 & **0.1323& **0.6933 	&**0.3025	& **0.1889 &	**0.2015 &**0.7780  \\
& & $\text{TaFR}_{GRV}$ & 0.2440 & 	0.1923& 0.1158&  0.6315&	0.3140& 	0.2149 & 0.1598& 0.7037 \\
\cmidrule(lr){4-7} \cmidrule(lr){8-11}
 & \multirow{3}*{GRU4Rec} & backbone
&0.2892&	0.2105& 0.0150 &0.2079 	&0.3885&	0.2425& 0.0169 & 0.2612\\
 & &$\text{TaFR}_{time}$& 0.2940&	0.2124&	0.0113& **0.1884 &0.3912&	0.2437 &0.0169& **0.2444 \\ 
& & $\text{TaFR}_{GRV}$ & **0.3129&	**0.2328& *0.0245& **0.3021 &**0.4111	&**0.2645 & **0.0357 & **0.3782\\
\cmidrule(lr){4-7} \cmidrule(lr){8-11}
& \multirow{3}*{TiSASRec} & backbone
&\textbf{0.3903} & 	\textbf{0.3298} &  \underline{0.4211}&	\underline{0.7876} & \textbf{0.4713} &	\textbf{0.3559}  & \underline{0.4862} & \underline{0.8356} \\
 & &$\text{TaFR}_{time}$ 
 & **0.3018 &	**0.2316 & **0.2005 & **0.3816  & **0.3932& 	**0.2611  &**0.2807& **0.5017  \\
 &  & $\text{TaFR}_{GRV}$  & \underline{0.3893} 	& \underline{0.3278} & \textbf{ 0.4236}& \textbf{0.7899 }&\underline{0.4706} &\underline{0.3540}  & **\textbf{0.5013}  & \textbf{0.8373} \\
\bottomrule
\end{tabular}
}
\end{table*}

\subsubsection{\textbf{Evaluation Metrics}}
\ 
\newline
We evaluate the accuracy of recommendations and the fairness degree of exposure mechanisms simultaneously. For recommendation accuracy, we use standard metrics: hit rate at k~(HR@k) and normalized discounted cumulative gain at k~(NDCG@k). For exposure fairness, we report the newly uploaded items' coverage~(N\_Cov@k), measured as \#exposed\_new\_items\_in\_topK\_lists/\#new\_items, and also show the overall item coverage~(Cov@k) in Section~\ref{sec:exp_fairness_analysis}. In MIND, we mark the 20\% of items with the latest upload time in the training set as new items. For Kuai, we label the newly uploaded 10\% of items as Kuai is less sparse than MIND. 
For each setting, we run experiments with random seeds $0\sim4$ and  
report the average results with significant tests compared with the backbone model.

\subsection{Overall Performance}
\label{sec:exp_overallPf}
The overall performance of the Timeliness-aware Fair Recommendation Framework is reported in Table \ref{table:recommendation}. 
We evaluate TaFR' recommendation quality and time-sensitive exposure fairness on the backbone model only, $\text{TaFR}_{time}$ and $\text{TaFR}_{GRV}$. 

(1) In general, TaFR achieves 
steady improvements in recommendation accuracy and time-sensitive exposure fairness simultaneously across different types of backbones and datasets compared with using backbone models only. This verifies the validity of explicitly modeling item timeliness distribution.
Compared with integrating upload time baseline~($\text{TaFR}_{time}$), TaFR with our proposed GRV module~($\text{TaFR}_{GRV}$) improves new items' coverage with steady recommendation quality. This proves the effectiveness of GRV module designs.

(2) For MIND dataset, the three types of backbones show an upward trend in both recommendation accuracy and new items' coverage on the top 10 results. Integrating Global Residual Value reduces the gap. When adopting the basic model NeuMF as the backbone, TaFR achieves comprehensive better results. Note that TaFR brings in lifts of recommendation quality and new items' exposure coverage based on the time-aware recommendation algorithm, TiSASRec. This further points out that the item-side timeliness distribution is an important modeling direction different and independent of user-side sequence modeling.

(3) In Kuai, the backbone model only, TaFR integrated with upload time and TaFR with GRV module all maintain increasing patterns when the backbone model is used in order with the basic, sequential, and time-aware methods. At the same time, TaFR with the GRV module consistently surpasses the +time baseline in both time-sensitive item exposure fairness and recommendation quality.

(4) Results on the two datasets show relatively different trends and this may relate to the different sparsity and timeliness patterns.
Recommendation performances and item coverages in MIND are relatively low. This pattern is consistent with the interaction sparsity and fast iteration characteristics of the MIND news dataset. Improvements in both recommendation quality and fairness of TaFR with GRV module integrated are more significant in MIND. This is probably because the items in MIND are uploaded on different days and for Kuai, items are uploaded within 24 hours.



\subsection{Impact on Time-sensitive Exposure Fairness}
\label{sec:exp_fairness_analysis}
\begin{figure}[htbp]
    \centering
    \subfigure[Top 5 results of Kuai, GRU4Rec]{
        \includegraphics[width=0.47\linewidth]{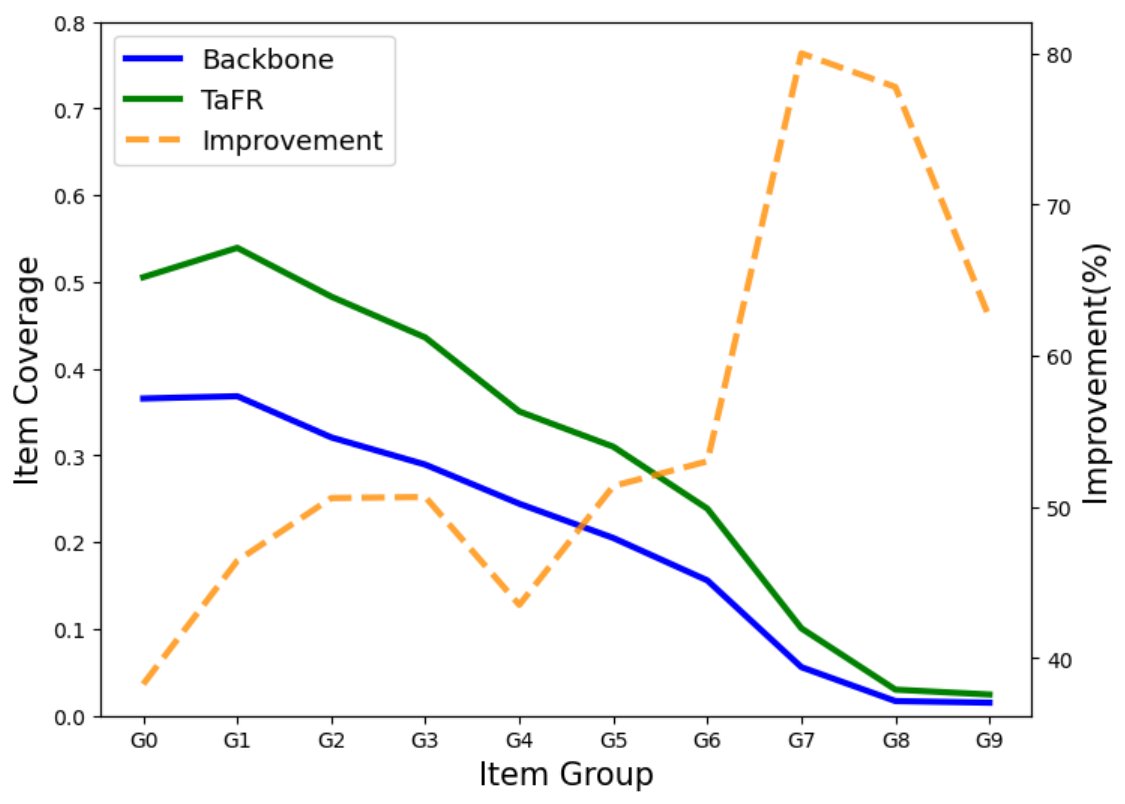}
        \label{fig:exp_kuai_5}
    }
    \subfigure[Top 10 results of Kuai, GRU4Rec]{
        \includegraphics[width=0.47\linewidth]{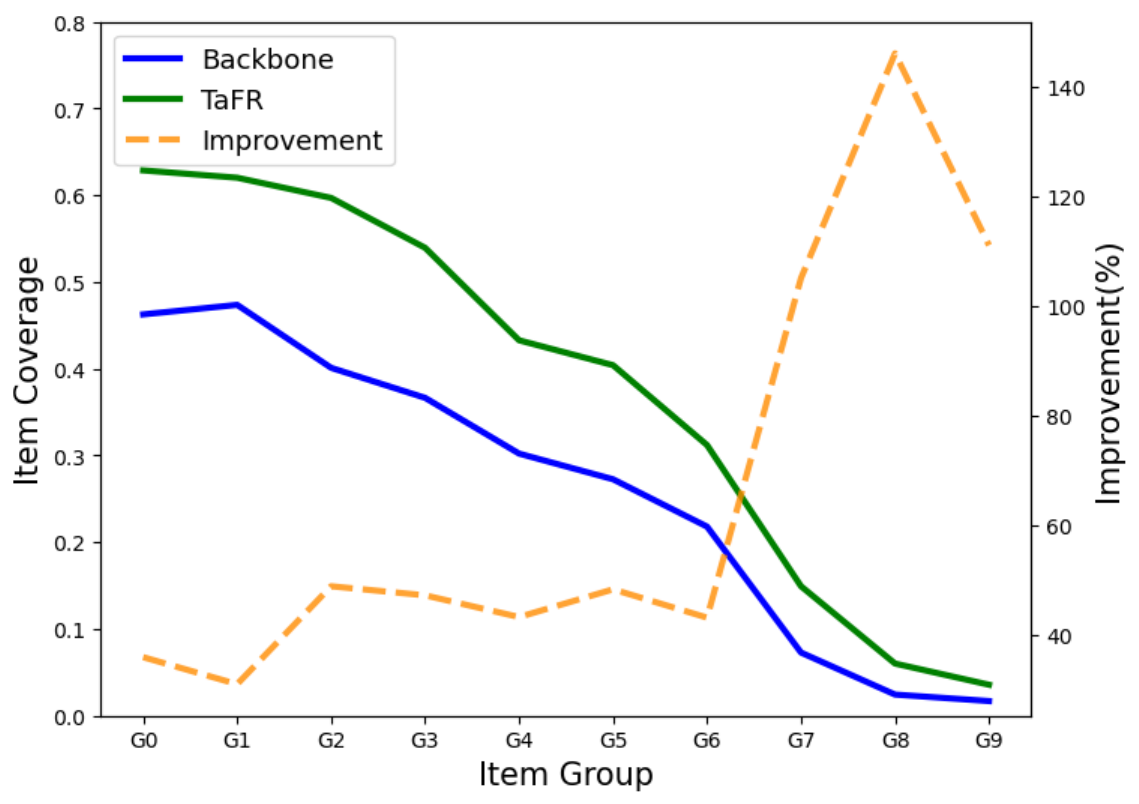}
        \label{fig:exp_kuai_10}
    }
\subfigure[Top 5 results of MIND, TiSASRec]{
        \includegraphics[width=0.47\linewidth]{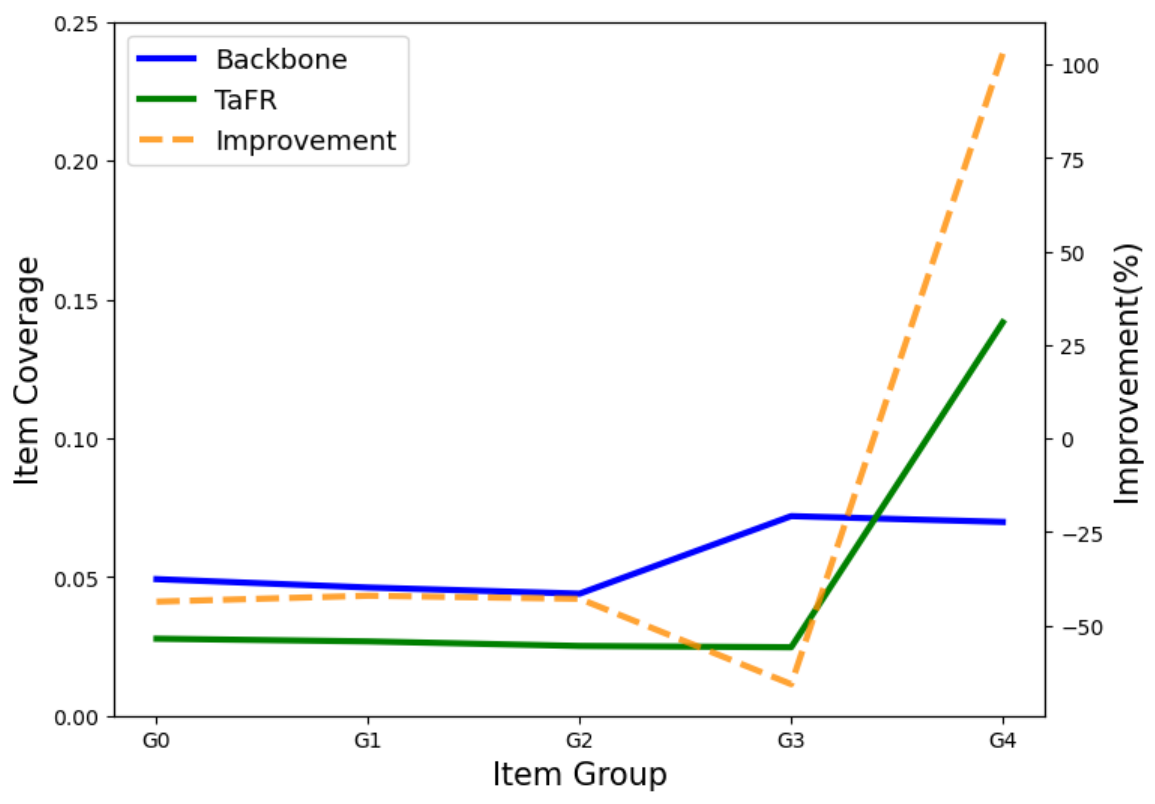}
        \label{fig:exp_mind_5}
    }
    \subfigure[Top 10 results of MIND, TiSASRec]{
        \includegraphics[width=0.47\linewidth]{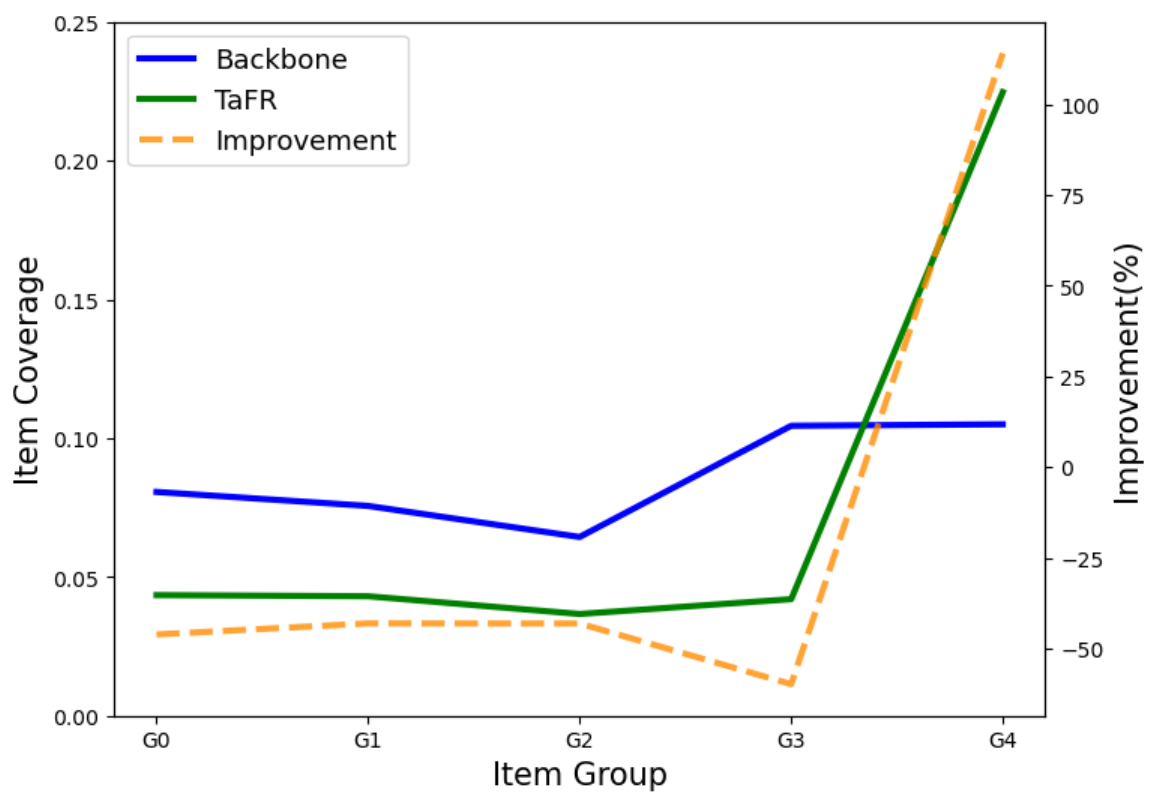}
        \label{fig:exp_mind_10}
    }
    \caption{Exposed item coverage among items with different upload times. Higher group numbers represent later upload times. The backbone model only, $\text{TaFR}_{GRV}$, and its percentage lift based on the backbone are shown, validating that TaFR boosts exposures opportunities.}
    \label{fig:exp_fairness}
\end{figure}
We further evaluate TaFR's modification on exposure between items uploaded at different times. Figure~\ref{fig:exp_fairness} groups all the items into 10 groups in Kuai and 5 groups in MIND evenly according to upload times and presents each group's exposed item coverage when recommend with the backbone~(in blue) or TaFR~(in green), and the TaFR's lift percentage based on backbone~(in orange). 
Figure~\ref{fig:exp_kuai_5},\ref{fig:exp_kuai_10} adopt GRU4Rec as the backbone and plot the Top 5 or 10 recommendation results in Kuai.
As the group number increase, items are uploaded later and the exposed item coverage decreases significantly.
TaFR constantly improves each group's item coverage in recommendation lists compared with the backbone and the boosted percentages are much higher for later uploaded items.
Figure~\ref{fig:exp_mind_5},\ref{fig:exp_mind_10} select TiSASRec as the backbone and is tested in MIND. TaFR greatly improves coverage of the last uploaded group, further enhancing the exposure of new items in the feed system.

\subsection{Evaluation on Global Residual Value Module}
\label{sec:exp_GRV}
\begin{figure}[htbp]
    \centering
    \subfigure[play rate]{
        \includegraphics[width=0.475\linewidth]{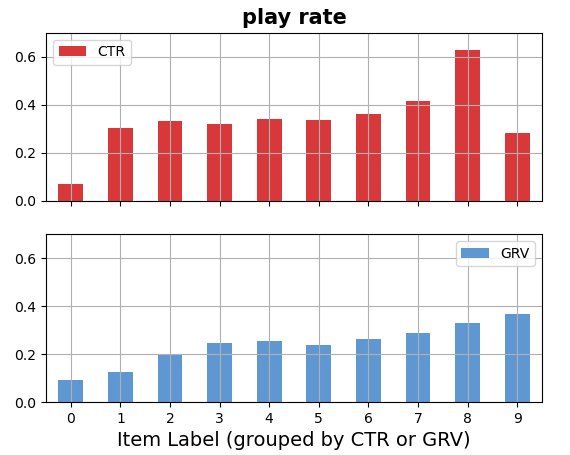}
        \label{feedback on Play Rate}
    }
    \subfigure[comment rate]{
	\includegraphics[width=0.475\linewidth]{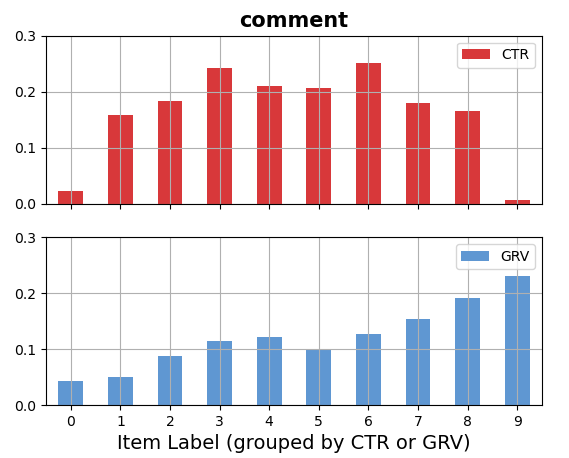}
        \label{feedback on Comment}
    }
    \caption{Items are divided into 10 groups based on historical CTR~(upper fig, colored in red) or GRV~(lower fig, colored in blue), respectively. We present each group's performances on two user feedback dimensions, the play rate for (a) and comment for (b). GRV shows more accurate modeling capabilities compared with history CTR.}
    \label{fig:GRV_examinations}
\end{figure}

This section examines the GRV module's ability in modeling the global timeliness of items. We carry out an analytical experiment for verification that the module produced GRV matches items' timeliness characteristics in future times. Following PageRank's bucket evaluation methods~\cite{davison2006propagating}, we divide items in the Kuai dataset into 10 groups based on historical CTR (baseline) and GRV module's output, $GRV_i(T_{i,0}+48)$, at $T_{i,0}$+48 hours and evaluate the group-level performance on user's feedback in play rate (playtime divided by video duration) and comment rate, which are distinguished from the input CTR.
Figure~\ref{fig:GRV_examinations} presents the results. We can see that blue bars show a monotonically increasing trend in diverse user feedback, while red bars have no apparent distinguishing abilities between groups. This result points to the modeling capabilities of the GRV module in TaFR and the possible problems in using linear CTR weighting methods as item timeliness modeling. 

%% file: sections/S6-conclusion.tex
\section{Conclusion}
In the era of data explosion, recommendation systems play a crucial role in people's access to information. However, information systems often seek to maximize delivery accuracy for higher user satisfaction and duration times, leaving relatively less attention on designing item-side fair exposure mechanisms, resulting in resource allocation issues such as the Snowball Effect.

In this paper, we first investigate the exposure situation between items uploaded at different times and name it as the \textit{time-sensitive exposure} issues.
Analyses point out the Snowball Effect may be caused by the time-sensitive unfair exposure and items in feed scenarios have diverse timeliness patterns.
Then, we propose to explicitly model the item-level customized timeliness distribution, namely Global Residual Value, and introduce the designed GRV module into recommendations with the Timeliness-aware Fair Recommendation Framework, TaFR, aiming to alleviate time-sensitive unfair exposures with minimal negative or positive impacts on recommendation accuracy. Abundant experiments are conducted with three types of backbone models on the two datasets, validating the ability of the TaFR for simultaneous improvements in both time-sensitive exposure fairness and recommendation quality.